# Fabrication and Characterization of Two-Dimensional Photonic Crystal Microcavities in Nanocrystalline Diamond

C. F. Wang,[a),b)] R. Hanson,[a)] D. D. Awschalom,[a)c)] E. L. Hu[c),d)]

University of California, Santa Barbara, California 93106

T. Feygelson, J. Yang, J. E. Butler

Gas/Surface Dynamics Section, Naval Research Laboratory, Washington DC 20375

## ABSTRACT

.

Diamond-based photonic devices offer exceptional opportunity to study cavity QED at room temperature. Here we report fabrication and optical characterization of high quality photonic crystal (PC) microcavities based on nanocrystalline diamond. Fundamental modes near the emission wavelength of negatively charged nitrogen-vacancy (N-V) centers (637 nm) with quality factors (Qs) as high as 585 were observed. Three-dimensional Finite-Difference Time-Domain (FDTD) simulations were carried out and had excellent agreement with experimental results in the values of the mode frequencies. Polarization measurements of the modes were characterized; their anomalous behavior provides important insights to scattering loss in these structures.

___________________________

[a)] Department of Physics

[b)] Electronic mail : chioufu@physics.ucsb.edu

[c)] Department of Electrical and Computer Engineering

d) Materials Department

Negatively charged nitrogen-vacancy (N-V) centers in diamond have attracted much attention recently due to their unique properties,[1] such as very long spin lifetimes at room temperature,[2] and their suitability for single photon sources.[3-4] When such sources are matched by the formation of high quality cavities, opportunities arise for enhancement and control of the optical transitions associated with N-V centers. We have previously observed whispering-gallery modes of nanocrystalline diamond microdisks with diameters of 10 and 15 μm.[5] To enhance light-matter interactions, smaller cavities with more highly concentrated electric fields are required. High quality factor (Q) photonic crystal (PC) microcavities are promising structures for this purpose, and several articles have discussed the possibilities of using diamond based photonic crystal cavities for quantum electrodynamics applications.[6,7] Although there have been previous reports of photonic crystal fabrication in a nanocrystalline diamond film,[8] there has been no description of the optical performance of diamond photonic crystals. This paper describes the fabrication and optical characterization of L7 photonic crystal cavities formed from nanocrystalline diamond. Although limited by surface roughness and grain boundaries in the material, the quality factors of the fundamental modes are observed to be as high as 585. Polarization measurements of the emitted photons are also discussed in detail.

A nanocrystalline diamond film was grown on 1μm-$SiO_2$/Si substrates by a modified commercial (Astex) 1.5 kW microwave plasma-enhanced CVD system using $CH_4$ and $H_2$ reactant gases. We note that the thickness of diamond film on a 3 inch wafer varies from 100 to 350 nm. Samples with thicknesses between 140 to 160 nm were used in this study. The grain size of the diamond is about 10 to 30 nm. $SiO_2$, patterned by electron beam lithography and reactive ion etching (RIE), served as a mask for the subsequent oxygen RIE transfer of the patterns into diamond. SEM images of suspended L7 photonic crystal cavities are shown in Fig.1. The cavity consists of seven missing holes along the Γ-K direction in a triangular lattice PC structure, with a lattice constant (a) of 240 nm and radius of the air holes (r) of about 80 nm The cavity was designed to form modes resonant with the zero phonon line (ZPL) of negatively charged N-V centers (about 637 nm). Fig.1(c) shows a cross sectional SEM of the air holes. The sidewall is fairly vertical, with a slight tilt of 3 degrees. This taper may result from a combination of the mechanism of etching, and continuous erosion of the $SiO_2$ mask during the diamond etching.

Micro-Photoluminescence (μ-PL) measurements were used to characterize the optical properties of the photonic crystal cavities. A continuous wave (CW) laser at 532 nm with 15 mW power was used to excite the devices. An objective lens with a numerical aperture of 0.55 was used to focus the laser onto the samples. The PL of the devices was collected by the same objective, passed through a monochromator, and detected by a liquid-nitrogen-cooled CCD. No emission from N-V centers was observed in our material. Our samples demonstrate a broad luminescence, as seems typical for nanocrystalline diamond films, and attributed to defect states within the grains.[9] The PL spectrum of a PC cavity is shown as the black curve in Fig.2. Five modes are clearly observed. The profiles of the |Hz| and |E| components of the four lowest-energy modes, obtained through three-dimensional Finite-Difference Time-Domain (FDTD) simulations,[10] are also shown in Fig. 2. As classified by S.-H Kim et al.,[11] the three lowest-energy modes, at 631, 626, and 610 nm, are even modes, and can be named as e1, e2, and e3 respectively; whereas the mode at 592 nm is an odd mode, named o1. To accommodate the constraints of the computation, different symmetric boundary conditions were applied to calculate the expected frequencies of the lowest energy modes for this cavity and the resulting modes are displayed as the colored curves in Fig 2. The parameters used in the simulation were: a = 240 nm, r = 80 nm, thickness = 150 nm, and refractive index (n) = 2.33.[12] When the simulated spectra were shifted 12 nm to longer wavelengths, we were able to achieve excellent matches of the four lowest energy modes to the experimental data. The discrepancy between simulation and experimental data arises from the variation in thickness of the nanocrystalline diamond films. Simulations also allowed us to calculate expected values of Q, and both simulated and experimental values (averaged from 15 devices) of Q are listed in Table 1. The standard deviation reflects the variation in Q values for each particular mode transition. The highest Q of the fundamental mode was measured to be 585.

Due to the symmetry of the structure, emissions from L7 cavities exhibit linearly polarized characteristics in far-field detection.[11, 13] We used these polarization characteristics to provide further insights to the values of Q measured for these devices. The polarization of each mode can be understood from an inspection of their in-plane field components. Ex and Ey components of both e1 and o1 are shown in Fig.3, together with the coordinate system we use. For mode e1, as well as all the other even modes, $E_x$ at y > 0 is the same as $E_x$ at y < 0, with a phase difference of 180 degrees (denoted antisymmetric about y=0). For z >> 0 these two fields would cancel out, resulting in zero $E_x$ in the far field. In contrast, the $E_y$ component is symmetric about y=0, and the components would add, rather than cancel in the far field. Therefore, all the even modes in the far field show a net y-polarization. For the o1 mode, the cancellation effects take place for the $E_y$ components; resulting in x-polarized light in the far field. To confirm this experimentally, a linear polarizer was placed in front of the monochromator. In figure 4(a), the black curve is the PL spectrum taken without a polarizer. The red (blue) curve denotes the PL taken with a polarizer oriented along the y (x) directions. Under these conditions, one would expect that all even modes (having y-polarization) appear only in the red spectrum, and all odd modes (having x-polarization) appear in the blue spectrum. The expected polarization behavior is shown in Figure 4(a). However, 10 out of 15 cavities measured showed the polarization characteristics shown in Figure 4(b), where the e1 mode has components of polarization in both the x and y-directions. Under the condition of using an x-polarizer, Figure 4(c) shows the mode intensities to background emission ratio for mode e1 of 15 devices plotted in a histogram. All 15 devices displayed the same mode frequencies, with a similar range of Q values for each mode. Thus we believe that variations in fabrication (e.g. variations in lattice constant, hole radius, sidewall angle) would not account for the considerable spread in the polarization of the e1 mode observed among the different devices. Rather, the possible reasons underlying the variations in polarization may arise from local variations in the material.

As is evident from the data shown in Table 1, the largest difference between the expected, simulated value of Q and the experimentally observed value is for the e1 mode: simulations predict a Q of 9100, while experiments yielded a value of 585. For all other modes, the experimentally obtained Q values were comparable to the simulated values. Similarly, we observed the largest polarization anomaly for the e1 mode, and to a much smaller extent, the e2 mode. Q values are deduced from the FDTD simulations through calculations of the radiative lifetime of the photons within the photonic crystal. These calculations assume 'perfect materials' with a uniform refractive index, and do not account for defects in the material, nanocrystalline structure and grain boundaries, all of which may give rise to different mechanisms of optical loss in the material. The polarization anomaly observed for the e1 mode might result from scattering from grain boundaries within the diamond and subsequent emission from a non-smooth cavity surface. If scattering has such a large effect on the observed polarization of the e1 mode, it may likewise be linked to principal mechanisms of optical loss, therefore reducing the observed Q for that mode. In the case of the o1, e2 and e3 modes, simulations suggest a much lower intrinsic Q and hence shorter radiative lifetimes. These lifetimes may be comparable to, or shorter than typical scattering lifetimes. Hence the effects on polarization are less easily observed. In addition, such scattering does not provide the principal limitation to Q (as is true for the e1 mode), and the discrepancy between simulated and observed Qs are not as great as for the e1 mode. These results suggest that photonic crystal cavities fabricated from single crystal diamond should demonstrate far higher values of Q.

In summary, nanocrystalline diamond based photonic crystal microcavities were fabricated and their optical properties were characterized. Three-dimensional FDTD simulations were used to identify the observed modes in PL spectra. The Qs of the fundamental mode, e1, were measured to be as high as 585. The polarization measurements suggest that scattering loss, associated with the nanocrystalline nature of the material, can be the main limiting factor on Q. We expect that higher performance can be achieved from photonic crystal cavities fabricated from single crystal diamond.

C. F. Wang would like to thank Min-Kyo Seo, Tsung-Li Liu, Yong-Seok Choi, and Kevin Hennessy for useful discussions. This work is supported by the DMEA under the Center for Nanoscience Innovation for Defense (CNID).

# Figure Captions

Figure 1. (a) A SEM image of a suspended L7 cavity, with a = 240 nm and r ≈ 80 nm. (b) An enlarged picture of air holes. (c) A cross sectional image of the air holes. The thickness of the membrane is about 160 nm. The sidewall of the holes is tilted by 3 degree.

Figure 2. (color online) (a) PL spectrum of a L7 cavity is shown in the black line, with the wavelength noted at the bottom. Simulated data is shown in colored lines, with the wavelength noted at the top. Curves with different colors correspond to different symmetric boundary conditions used in the simulation. Mode profiles of |Hz| and |E| for four modes, e1, e2, e3, and o1 are shown at the top.

Figure 3. (color online) Mode profiles of Ex and Ey components of e1 and o1. Coordinate systems are defined in the bottom figure.

Figure 4. (color online) (a)(b) For two different devices, PL spectra with (without) a linear polarizer are shown as colored (black) curves. The orientations of the polarizer are vertical and horizontal for red and blue curves, respectively. (c) A histogram of mode intensity (for mode e1) to background ratio, with an x-polarizer, of 15 devices.

**Table** 1. Simulated and experimental quality factors for four lowest-energy modes. Qs$_{experimental}$ were averaged from 15 devices.

| Modes | e1 | e2 | e3 | o1 |
|---|---|---|---|---|
| Q$_{simulation}$ | 9100 | 830 | 230 | 480 |
| Q$_{experimental}$ | 525 ± 33 | 373 ± 29 | 163 ± 21 | 335 ± 32 |

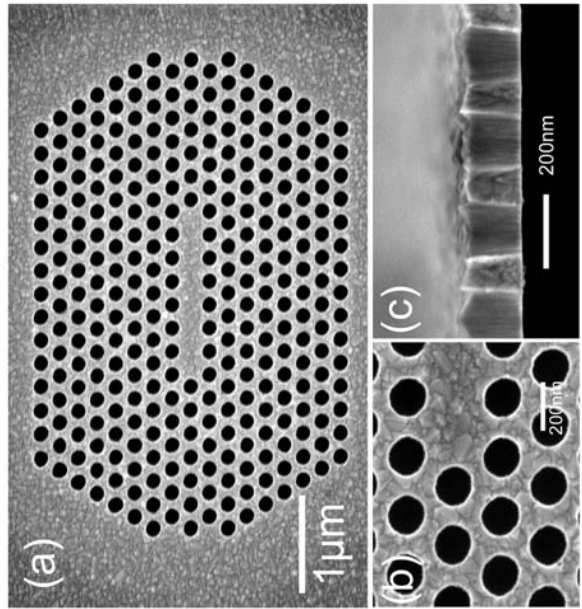

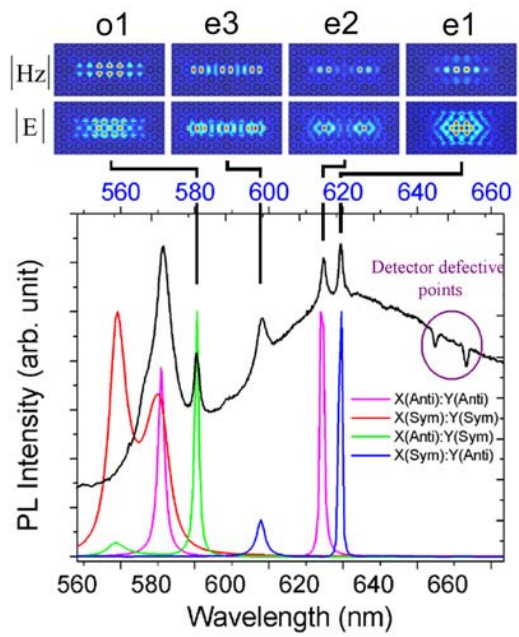

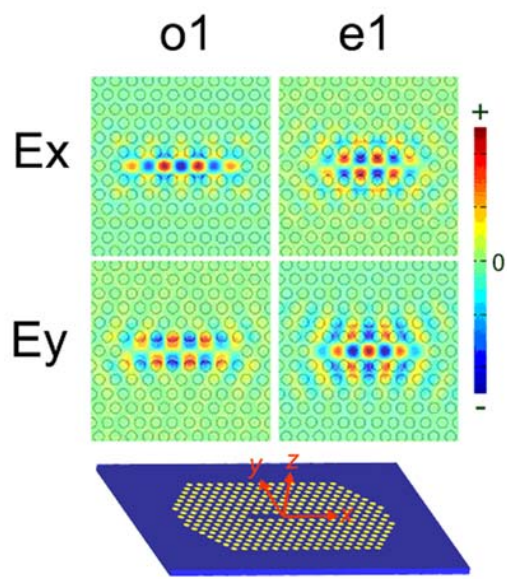

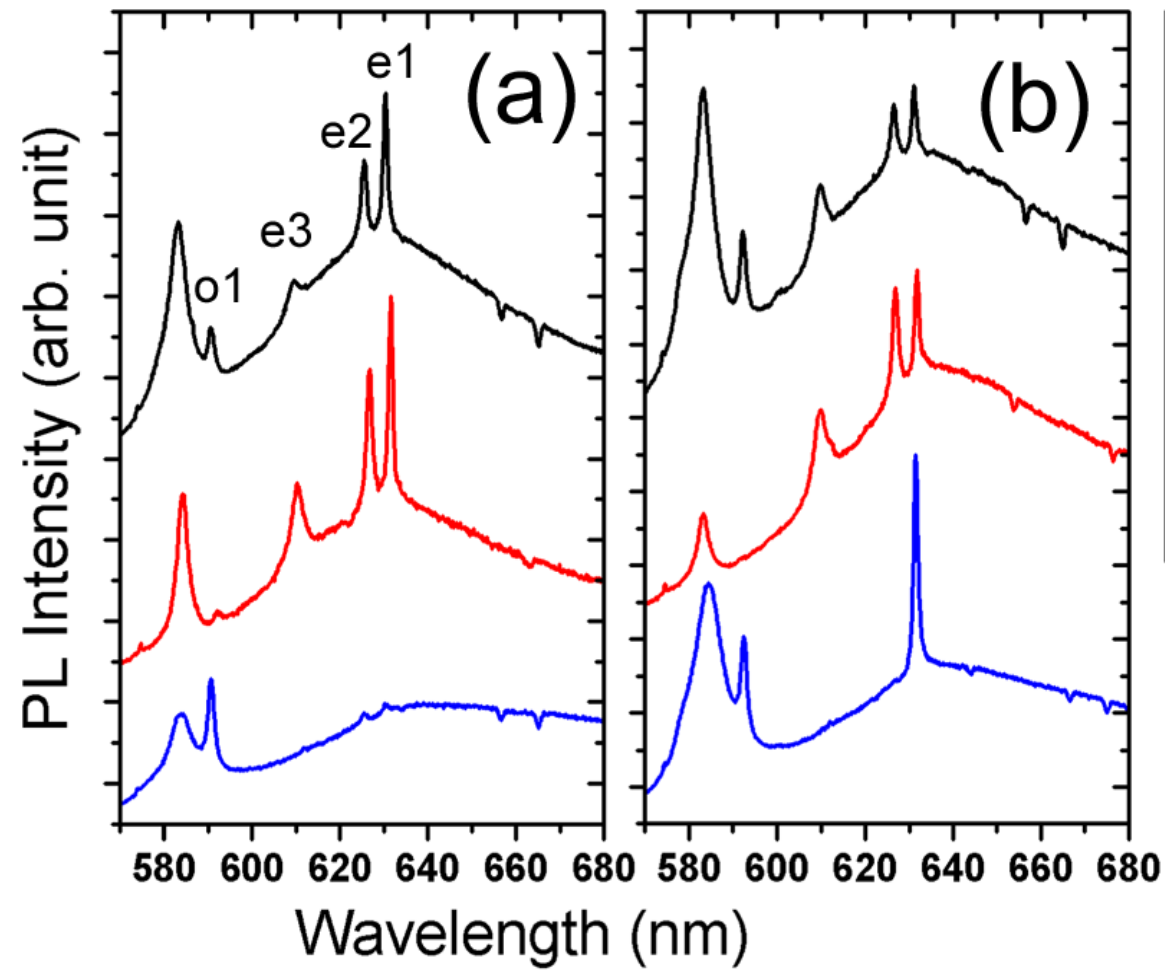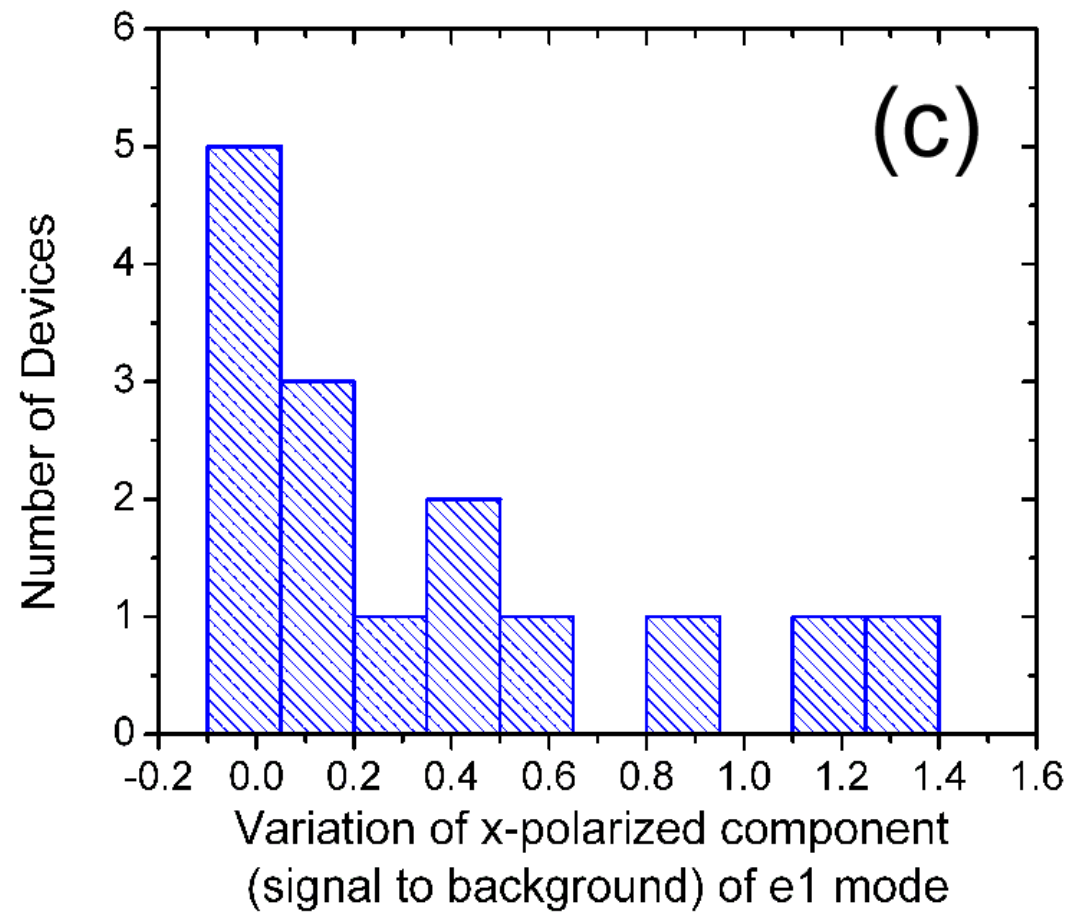